\documentclass[aps,prl,twocolumn,showpacs,superscriptaddress]{revtex4}
\usepackage{amsmath}
\usepackage{amsfonts}
\usepackage{amssymb}
\usepackage{epsfig}
\newcommand{\be}{\begin{equation}}
\newcommand{\ee}{\end{equation}}

\begin{document}

\title{Heterogeneous diffusion in a reversible gel}

\author{Pablo I. Hurtado}
\affiliation{Laboratoire des Collo\"{\i}des, Verres et Nanomat\'eriaux, 
UMR 5587, Universit\'e Montpellier II and CNRS, 
Montpellier 34095, France}
\affiliation{Departamento de Electromagnetismo y F\'{\i}sica de la Materia, 
and Instituto Carlos I \\
de F\'{\i}sica Te\'orica y Computacional, Universidad de Granada, 
Granada 18071, Spain}
\author{Ludovic Berthier}
\author{Walter Kob}
\affiliation{Laboratoire des Collo\"{\i}des, Verres et Nanomat\'eriaux, 
UMR 5587, Universit\'e Montpellier II and CNRS, 
Montpellier 34095, France}

\date{\today}

\begin{abstract}
We introduce a microscopically realistic model of a 
physical gel and use computer simulations to study its 
static and dynamic properties at thermal equilibrium.
The phase diagram comprises a sol phase, 
a coexistence region ending at a critical point,
a gelation line determined by geometric percolation, and an 
equilibrium gel phase unrelated to phase
separation. The global structure of the gel is homogeneous, 
but the stress is only supported by a fractal network.
The gel dynamics is highly heterogeneous and 
we propose a theoretical model to quantitatively 
describe dynamic heterogeneity in gels. 
We elucidate several differences between the dynamics of 
gels and that of glass-formers.
\end{abstract}

\pacs{61.43.Bn, 82.70.Gg, 61.20.Lc}


\maketitle

Although gels are commonly used in everyday life
they continue to offer fundamental challenges to research.
Their physics is determined by a wide window
of lengthscales, from the molecular size of particles
in the solvent to macroscopic structures, 
and by a similarly broad range of timescales: 
Gels are ``complex'' fluids~\cite{larson}.
Of particular interest are physical
gels which are typically made of molecules forming a 
stress-sustaining network, with links that have a finite lifetime, 
as opposed to chemical gels where junctions are permanent and 
properties follow directly from geometry. 
The transient character of the network in physical gels results in
a complex interplay between structure and dynamics, 
leading to non-trivial flow properties.
Here we propose a model of a reversible physical gel which is
microscopically realistic (we are in fact inspired by 
one particular material), and specifically design
a hybrid Monte Carlo / molecular dynamics numerical approach 
to successfully bridge the gap between 
microscopic details and macroscopic observations, 
while offering deep insight on the nature of physical gels. 

Inspired by recent experimental work on gelation, a variety of ``minimal'' 
models have recently been studied to elucidate the connection between 
gelation and seemingly related phenomena: Geometric 
percolation~\cite{napoli,jullien}, 
glass transition~\cite{fs1,puertas,ema}, phase 
separation~\cite{fs2,dave}. 
Detailed experiments performed with colloidal particles with tunable 
interactions~\cite{colloids} revealed that a non-trivial interplay between  
phase separation and kinetic arrest may produce gel-like structures.
Associating polymers constitute 
another well-studied example of reversible gels~\cite{larson}. 
In that case, gels can be obtained far from  phase separation, 
producing viscoelastic materials with highly 
non-linear rheological properties that are not well 
understood~\cite{polymers,porte}. 
In many cases, a close similarity between gelation and 
glass formation is reported~\cite{larson}.
We explain below this similarity but discuss also important 
differences.

\begin{figure}[b]
\psfig{file=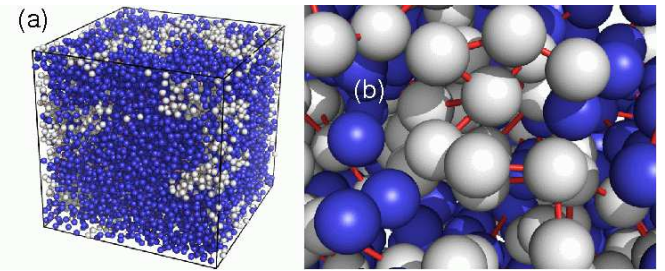,width=8.5cm}
\caption{\label{zoom}
(a) Structure of the physical gel
for $\phi=0.2$, $R=2$ and $N=10^4$. A percolating (light gray) cluster
of droplets connected by telechelic (red) polymers, through 
which the remaining (dark blue) droplets can diffuse, 
see zoom in (b).}
\end{figure}

Our model is inspired by a material described in
Ref.~\cite{porte}. 
It is a microemulsion of stable and monodisperse oil droplets in water mixed 
with telechelic polymers, i.e. long hydrophilic chains 
with hydrophobic end caps.
A polymer can form a loop around a single droplet, or, 
more interestingly, a bridge between two droplets. 
Figure~\ref{zoom} is a snapshot taken from our simulations showing 
droplets and bridging polymers. 
For sufficiently high polymer concentrations, a percolating network 
can be formed (shown in light gray) and the system becomes a soft solid. 
However, thermally activated extraction of the 
hydrophobic heads leads to a slow reorganization
of the network structure, and the material eventually
flows at long times~\cite{tanaka}.
This material is interesting because 
functionality, lifetime of the bonds, 
volume fraction, strength of the networks can all 
be adjusted independently, which is not 
always possible in attractive colloids~\cite{colloids}, or in previous
model systems, unless specific adhoc assumptions are 
made~\cite{fs1,puertas,ema,fs2}. 
Modeling such a complicated self-assembly is a  challenge because 
of the wide range of scales involved. 
In our model we neglect the solvent and 
include polymers and droplets as the elementary objects.
Moreover, since the internal dynamics of the polymers 
is much faster than the gel 
dynamics, we coarse-grain the polymer description and only 
retain their effect as links inducing an effective 
entropic interaction between the two droplets they connect.
Coarse-graining is a crucial step for efficient large scale 
simulations, not used in previous models~\cite{tanakasimu}.

We consider an assembly of $N$ droplets
of diameter $\sigma$ and mass $m$, interacting, in the absence
of polymers, with a pair potential typical of soft spheres, 
$V_{1}(r_{ij}) = \epsilon_1 \left({\sigma}/{r_{ij}} \right)^{14}$,
where $r_{ij}$ is the distance between
droplets $i$ and $j$, and $\epsilon_1$ an energy scale.
The potential is cut off and regularized at a finite distance,
$2.5 \, \sigma$.
In addition, $N_{\text{p}}$ polymers of maximal extension $\ell$ 
can form bridges between droplets, or loops. Polymer loops 
have an energy cost $\epsilon_{0}$, but no effect on droplet dynamics. 
On the other hand, bridging polymers induce an entropic 
attraction between connected droplets, which we model using
the classic FENE form, 
$V_2(r_{ij}) = - \epsilon_2 \ln \left[ 1-{(r_{ij}-\sigma)^2}/{\ell^2} 
\right]$, 
so that polymers act as springs at small elongation, but
cannot become longer than $\ell$. 
A configuration is specified by 
the droplets positions and velocities, $\{{\bf r}_i(t), 
{\bf v}_i(t)\}$, and by the polymer $N \times N$ connectivity 
matrix, $\{C_{ij}\}$,
where $C_{ij}$ is the number of polymers connecting droplets
$i$ and $j$. Summarizing, the Hamiltonian is thus
\be 
{\cal H} = 
\sum_{i=1}^N \Big( \frac{m}{2} {\bf v}_i^2+  C_{ii} \epsilon_{0} + 
\sum_{j>i} \left[ V_1(r_{ij}) + C_{ij} V_2(r_{ij}) \right] \Big).
\ee

Simulation proceeds by solving Newton's 
equations for the droplets.
Lengthscales are given in units of $\sigma$, energy in units
of $\epsilon_1$, and times in units of $\sqrt{m \sigma^2/\epsilon_1}$.
We use the velocity Verlet algorithm with discretization $h=0.005$. 
Simultaneously, we use Monte Carlo dynamics to evolve polymers.
In an elementary move, a polymer is chosen at random, and
one of its end caps is moved to a randomly chosen neighboring droplet.
This proposed move is accepted with rate 
$\tau_{\text{link}}^{-1} \text{min}[1,\text{exp}(-\Delta V_2/T)]$, 
where $\Delta V_2$ is the potential energy change during the move, 
$T$ is the temperature, and 
$\tau_{\text{link}}$ controls the timescale for 
polymer rearrangements.
In experiments $\tau_{\text{link}}$ has an Arrhenius behavior
associated to the excitation cost for polymer extraction.
We set $\ell=3.5 \, \sigma$~\cite{porte}, $T=1$, $\epsilon_0=1$, 
and $\epsilon_2=50$. We found little influence
of $\epsilon_2$ on the phase diagram.  
The relevant control parameters are the droplet 
volume fraction, $\phi=\pi N / (6V)$, where $V$ is the volume,
the number of polymer heads per 
droplet, $R=2 N_{\text{p}}/N$, and $\tau_{\rm link}$, which
has no influence on static properties. 
We performed simulations for a wide range of parameters,
$R\in[0,18]$, $\phi\in [0.01,0.3]$, $\tau_{\rm link} \in [1,10^4]$,
$N \in [10^3,10^4]$.

\begin{figure}
\centerline{
\psfig{file=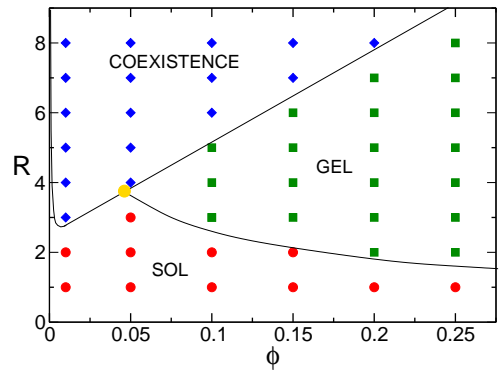,width=6.5cm}}
\caption{\label{phase} Phase diagram for a wide range
of volume fraction, $\phi$, and number of polymer heads per droplet, $R$.
Symbols are the investigated state points in 
the sol ($\bigcirc$), gel ($\Box$), and phase separated ($\diamond$) 
regions, the yellow point is the approximate location of the critical
point. Transition lines are sketched.} 
\end{figure}

The phase diagram, as obtained after 
a systematic exploration of the control parameter space,
is shown in Fig.~\ref{phase}.
Its topology is in good agreement with experiments 
in the range studied~\cite{porte}. 
In the low-$\phi$, low-$R$ region, the system resembles 
a dilute assembly of soft spheres, 
and has simple-liquid properties: This is the sol phase.
Increasing $R$ increases the effective attraction between droplets,
so that phase separation occurs at large $R$ between a low-$\phi$, low-$R$
phase and a large-$\phi$, large-$R$ phase~\cite{prlporte}. 
We detect phase coexistence from direct visualisation
and by measuring the static structure factor, 
$S(q) = \langle N^{-1} \sum_{jk} \exp [ i {\bf q} \cdot 
({\bf r}_j - {\bf r}_k)] \rangle$,
which exhibits the typical $q^{-4}$ behavior at small $q$. 
We observe both nucleation 
or spinodal regimes depending on the quench depth.
Interestingly we find that the kinetics of the phase separation is 
extremely slow, and that the obtained patterns
are very similar to the ones observed experimentally in
short-range attractive colloidal suspensions~\cite{colloids}.

For $\phi>0.05$, a broad gel region exists at thermal
equilibrium between the sol phase
at low-$R$ and phase separation at large-$R$, see Fig.~\ref{phase}. 
In the gel phase, a system-spanning cluster of polymer-connected 
droplets emerges, which endows the fluid with viscoelastic properties, 
see Fig.~\ref{zoom}.
The sol-gel transition coincides with geometric percolation
of polymer-connected clusters. Our gels are homogeneous,
i.e. $S(q)$ remains typical of a simple fluid, 
as seen in experiments~\cite{porte}.
However, the spanning cluster is highly fractal near percolation,
and becomes thicker deeper in the gel phase. 
At percolation we find a distribution of cluster sizes
$P(n)\sim n^{-\gamma}$, with $\gamma\approx 2.2$ 
compatible with random bond percolation, as seen in other 
systems~\cite{napoli,fs1,ema,fs2}.
Finally the structure of the system becomes nontrivial
when approaching the critical point, where $S(q)$ 
develops a power law behavior with an exponent close to -1.5
at small $q$ for $\phi_c \approx 0.05$, $R_c \approx 3.5$. 

\begin{figure}
\psfig{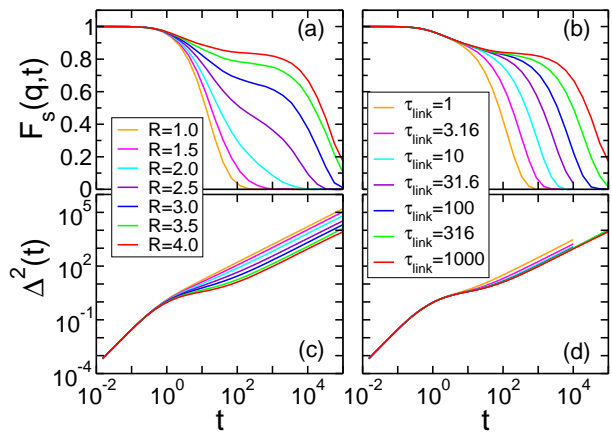}
\caption{Self-intermediate scattering function 
for $q=0.46$ (a, b) and mean-squared
displacement (c, d) for $N=10^3$, $\phi=0.2$. 
(a, c) present the dynamics for $\tau_{\rm link}=10^3$
and several values of $R$ across percolation ($R_{\rm p}\approx 1.85$).
Viscoelasticity continuously emerges at percolation. 
(b, d) are for $R=4$ and different values of $\tau_{\rm link}$, which
directly controls the long-time decay of $F_s(q,t)$, while 
$\Delta^2(t)$ remains essentially unchanged.}
\label{perco}
\end{figure}

We now show that the gel phase indeed behaves 
dynamically as a soft viscoelastic fluid.
We have investigated the dynamics 
by measuring the self-intermediate scattering function,
$F_s(q,t) = \langle N^{-1} \sum_{j} 
\exp[i {\bf q} \cdot ({\bf r}_j(t) - {\bf r}_j(0) 
)] \rangle$, and
the mean-squared displacement, 
$\Delta^2(t) = \langle N^{-1} \sum_j | {\bf r}_j(t)
- {\bf r}_j(0) |^2 \rangle$.
Figure~\ref{perco}-a shows the evolution of $F_s(q,t)$
from the sol to the gel phase. While relaxation is fast and exponential 
in the sol phase, a slow secondary relaxation emerges at percolation.
The final decay time varies little in the gel phase,
but the height of the plateau at intermediate times 
evolves dramatically. A similar
behavior is found for the coherent scattering function,
as in experiments~\cite{porte}. 
Physically the plateau reflects the thermal vibrations of an elastic
solid on intermediate timescale, while long-time decay reflects 
the flow of the system: The system is viscoelastic.
In Fig.~\ref{perco}-b we show that viscous flow is mostly controlled 
by $\tau_{\rm link}$, the rate for polymer 
extraction \cite{napoli,fs3}. Flow in this
system occurs when the percolating network slowly rearranges 
through polymer moves~\cite{tanaka}.
Therefore, gelation corresponds to the 
continuous emergence, for increasing
polymer concentration, of a plateau in dynamic functions, with
an almost constant relaxation timescale,
controlled by the polymer dynamics.
Gelation is thus qualitatively different from 
a glass transition where the plateau height
remains constant with a dramatic increase of 
relaxation timescales~\cite{debe}. Coincidence of gelation and 
percolation, put forward in~\cite{napoli} or
dispelled in~\cite{fs2}, happens
whenever long-lived bonds make cluster restructuration
very slow, but does not occur in systems where the bond lifetime
is short at percolation~\cite{fs1}.

Surprisingly, the mean-squared displacements
shown in Fig.~\ref{perco}-c-d appear as poor indicators of the dynamics.
The comparison between Figs.~\ref{perco}-b-d
is in fact quite striking. While the final relaxation timescale, $\tau$, 
in $F_s(q,t)$ scales roughly as $\tau_{\rm link}$, the diffusivity,
$D_s$,  extracted from the long-time behavior of 
$\Delta^2(t)\sim 6D_st$ is 
almost constant. This is reminiscent of the 
``decoupling'' phenomenon, or 
``breakdown'' of the Stokes-Einstein relation,
reported in supercooled fluids~\cite{debe}. While
``fractional'' breakdown is reported in liquids, $D_s \sim \tau^{-\zeta}$,
with $\zeta$ in the range 0.7-0.9 instead of the normal
value $\zeta=1$~\cite{mark}, we find here $\zeta \approx 0$, 
quite an extreme case of decoupling. Decoupling in gels
was reported in different systems~\cite{decoupling}.

\begin{figure}
\centerline{
\psfig{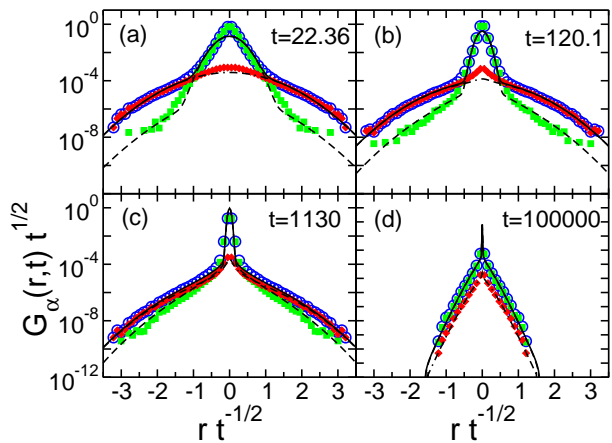}}
\caption
{Distribution of droplet displacements for $\phi=0.2$, $R=4$, 
$\tau_{\rm link}=100$ and different times for all droplets ($\bigcirc$), 
and its decomposition over droplets that
are initially free ($\diamond$) or arrested ($\Box$). 
Lines are from our model, Eq.~(\ref{FL}).}
\label{diff}
\end{figure}

In supercooled fluids, decoupling phenomena are commonly attributed 
to the existence of dynamic heterogeneity, that is, the existence
of non-trivial spatio-temporal distributions of mobilities.
The analogy is confirmed in Fig.~\ref{diff} where we show distributions 
of droplet displacements, $G_s(r,t) = \langle N^{-1} \sum_i 
\delta(|{\bf r}-{\bf r}_i(t)+{\bf r}_i(0)|) 
\rangle$. Clearly, $G_s$ exhibits a bimodal character suggesting 
coexistence of slow arrested droplets and fast diffusing droplets.
Qualitatively similar distributions were reported in 
gels~\cite{puertas,colloids} and glasses~\cite{mark,weeks}. 
Here, the snapshot in Fig.~\ref{zoom} suggests an obvious 
explanation for dynamic heterogeneity. At any given
time, a system-spanning cluster of droplets
which behaves as a solid on timescales smaller than $\tau_{\rm link}$
coexist with droplets which can more freely diffuse through this arrested
structure. We quantitatively confirm this interpretation in
Fig.~\ref{diff} where $G_s$ is decomposed over two families of 
droplets: $G_s = c_A G_A+ (1-c_A)G_M$, where $A$ ($M$) stands for 
droplets that are arrested (mobile) at time $t=0$, 
$c_A$ being the fraction of droplets belonging to the 
percolating cluster. While the central peak in $G_s$ is dominated by
$G_A$, the large ``non-Gaussian'' tails are dominated by $G_M$.

We now propose an analytic model to describe the 
dynamic heterogeneity of the gel, which incorporates 
the physical idea of a coexistence of 
a slow, percolating cluster of connected droplets
and fast, more freely diffusing droplets, with a
dynamic exchange between the two families set by polymer moves. 
Similar physical ideas were qualitatively discussed 
earlier~\cite{puertas2,decoupling,colloids}, but were 
however not exploited within a quantitative model.
We define $g_{\alpha}({\bf r},t)$, the 
probability that a droplet makes a displacement ${\bf r}$ in a time $t$ 
provided it belongs to family $\alpha$ during the 
whole time interval $[0,t]$, $\alpha=A$, $M$.
Let $p_{\alpha}(t)$ be the probability that a 
droplet in family $\alpha$ switches for the first
time to the complementary family, $\bar{\alpha}$, at time $t$,
and define $P_{\alpha}(t) \equiv 
\int_t^{\infty} p_{\alpha}(t') \text{d} t'$. Then we have
\be
G_{\alpha}({\bf r},t) =  P_{\alpha}(t) g_{\alpha}({\bf r},t) + 
\int_0^t dt' \left[
\Delta_{\alpha}({\bf r},t') \circ G_{\bar \alpha}({\bf r},t-t') \right],
\ee
where $\Delta_\alpha({\bf r},t) \equiv p_\alpha(t) g_\alpha({\bf r},t)$, 
and $\circ$ stands for spatial convolution.
The first term describes
droplets which persist in the same family between 0 and $t$,
while the second term captures family exchanges.
To keep the model simple, we assume 
the exchange dynamics to be homogeneous, 
with constant transition rates $\tau_\alpha^{-1}$, so
$p_\alpha(t)= \exp(-t/\tau_\alpha)/\tau_\alpha$. In addition, 
stationarity implies that $c_A / \tau_A = (1-c_A)/\tau_M$.
The model can be solved analytically in the Fourier-Laplace domain:
\begin{equation}
G_{\alpha}({\bf q},s)=\frac{\tau_{\alpha} \Delta_{\alpha}({\bf q},s) + 
\tau_{\bar \alpha} \Delta_{\alpha}({\bf q},s)
 \Delta_{\bar \alpha}({\bf q},s)}{1-\Delta_{\alpha}({\bf q},s) 
\Delta_{\bar \alpha}({\bf q},s)} \, .
\label{FL}
\end{equation}

To obtain analytic fits to the data we assume a Gaussian propagator for 
mobile droplets, $g_{{M}}({\bf r},t)=(4\pi D_{{M}}t)^{-3/2} 
\exp[- r^2/(4D_{{M}}t)]$, where $D_M$ is an effective diffusivity. 
We treat the droplets attached to the cluster as localized in 
a bounded region of space of linear size $a$, which  
reflects the thermal vibrations of the elastic solid,  
$g_{{A}}({\bf r})=(\pi a^2)^{-3/2} \exp[- r^2/a^2]$.
From Eq.~(\ref{FL}) we obtain a simple (but lengthy) analytic 
expression for the self-intermediate scattering function
$F_s(q,t)$ by inverse Laplace transform. We numerically  
invert the Fourier transform to get the displacement distributions.
The free parameters of the model are 
$\{ c_A, D_M, a, \tau_A \}$, but the first three 
parameters can be fixed by numerical observations.
The concentration of droplets connected to the 
percolating cluster, $c_{\rm A} \approx 0.94$, is directly measured. 
The value $D_M \approx 0.3$ is evaluated from an 
intermediate-time fit of $\Delta_M^2(t)$ restricted to 
initially mobile droplets.
The localization length, $a \approx 1.79$, is similarly estimated from 
the plateau in $\Delta_A^2(t)$ restricted to droplets 
initially belonging to the percolating cluster. 
To get the excellent fits shown in Fig.~\ref{diff}, 
we adjust $\tau_A =2 \cdot 10^4$, which coincides well
with the timescale at which $\Delta^2_A(t)$ becomes 
diffusive, a physically sound definition for 
the average exchange time.

We were able to fit data for a wide range 
of parameters and find that our model performs similarly well for other 
state points. From Eq.~(\ref{FL}), it is easy to predict
that $\tau \propto \tau_A$, while $D_s = (1-c_A)D_M + c_A a^2 / 
(4\tau_A)$, in quantitative agreement with the 
decoupling data reported in Fig.~\ref{perco}, leading to $\zeta = 0$ for 
large $\tau_{\rm link}$. 
The diffusion constant is in fact entirely 
dominated by those droplets which do not contribute to 
viscoelasticity, and is therefore a poor indicator of the gel dynamics.
These results show that dynamic heterogeneity in
gels can be stronger than in supercooled fluids, 
but its origin is also much simpler: The system structure is 
heterogeneous, Fig.~\ref{zoom}, while no such static structure 
exists in glasses.

Although motivated by a specific material, 
the new model for reversible gelation proposed in this 
work sheds light on the microscopic aspects of
gelation and the heterogeneous dynamics
of gel-forming systems. Moreover, our numerical findings
motivated a simple yet accurate analytic modeling 
of dynamic heterogeneity, which is generally applicable 
to gels. Although slow and heterogeneous dynamics are superficially
reminiscent of the physics of supercooled fluids,
we discussed several qualitative differences between gels and glasses.
 
\acknowledgments
Discussions with G. Porte, C. Ligoure, and S. Mora 
motivated this work. We thank T. Bickel and S. Kumar for discussions, 
DYGLAGEMEM, MEyC FIS2005-00791 and Universidad 
de Granada for financial support.

\end{document}